\documentclass[runningheads]{llncs}
\usepackage{graphicx}
\usepackage{hyperref}
\usepackage{booktabs} 
\usepackage{url}
\usepackage{mathrsfs}
\usepackage[mathscr]{eucal}
\usepackage[linesnumbered,ruled,vlined]{algorithm2e}
\usepackage{multirow}
\usepackage{subcaption}
\usepackage{amsmath}
\usepackage{graphicx}
\usepackage{array}
\usepackage{makecell}
\usepackage{setspace}
\usepackage{stfloats}
\usepackage{enumitem}
\usepackage{adjustbox}
\usepackage{wrapfig}

\usepackage{tikz}
\usetikzlibrary{arrows,positioning,automata,calc,shapes}
\usepackage{pgfplots}
\pgfplotsset{compat=newest, scaled z ticks=false} 
\pgfplotsset{plot coordinates/math parser=false}
\newlength\figureheight 
 \newlength\figurewidth

\newcommand{\squishlist}{
    \begin{list}{$\bullet$}
    { \setlength{\itemsep}{0pt}
        \setlength{\parsep}{1pt}
        \setlength{\topsep}{1pt}
        \setlength{\partopsep}{0pt}
        \setlength{\leftmargin}{1em} 
        \setlength{\labelwidth}{1em}
        \setlength{\labelsep}{0.5em}
    						 } }

\newcommand{\squishlisttwo}{
    \begin{list}{$\bullet$}
        { \setlength{\itemsep}{0pt}
            \setlength{\parsep}{0pt}
            \setlength{\topsep}{0pt}
            \setlength{\partopsep}{0pt}
            \setlength{\leftmargin}{2em}
            \setlength{\labelwidth}{1.5em}
            \setlength{\labelsep}{0.5em} } }

\newcommand{\squishend}{
    \end{list}  }

\makeatletter
\let\save@mathaccent\mathaccent
\newcommand*\if@single[3]{%
  \setbox0\hbox{${\mathaccent"0362{#1}}^H$}%
  \setbox2\hbox{${\mathaccent"0362{\kern0pt#1}}^H$}%
  \ifdim\ht0=\ht2 #3\else #2\fi
  }
\newcommand*\rel@kern[1]{\kern#1\dimexpr\macc@kerna}
\newcommand*\widebar[1]{\@ifnextchar^{{\wide@bar{#1}{0}}}{\wide@bar{#1}{1}}}
\newcommand*\wide@bar[2]{\if@single{#1}{\wide@bar@{#1}{#2}{1}}{\wide@bar@{#1}{#2}{2}}}
\newcommand*\wide@bar@[3]{%
  \begingroup
  \def\mathaccent##1##2{%
    \let\mathaccent\save@mathaccent
    \if#32 \let\macc@nucleus\first@char \fi
    \setbox\z@\hbox{$\macc@style{\macc@nucleus}_{}$}%
    \setbox\tw@\hbox{$\macc@style{\macc@nucleus}{}_{}$}%
    \dimen@\wd\tw@
    \advance\dimen@-\wd\z@
    \divide\dimen@ 3
    \@tempdima\wd\tw@
    \advance\@tempdima-\scriptspace
    \divide\@tempdima 10
    \advance\dimen@-\@tempdima
    \ifdim\dimen@>\z@ \dimen@0pt\fi
    \rel@kern{0.6}\kern-\dimen@
    \if#31
      \overline{\rel@kern{-0.6}\kern\dimen@\macc@nucleus\rel@kern{0.4}\kern\dimen@}%
      \advance\dimen@0.4\dimexpr\macc@kerna
      \let\final@kern#2%
      \ifdim\dimen@<\z@ \let\final@kern1\fi
      \if\final@kern1 \kern-\dimen@\fi
    \else
      \overline{\rel@kern{-0.6}\kern\dimen@#1}%
    \fi
  }%
  \macc@depth\@ne
  \let\math@bgroup\@empty \let\math@egroup\macc@set@skewchar
  \mathsurround\z@ \frozen@everymath{\mathgroup\macc@group\relax}%
  \macc@set@skewchar\relax
  \let\mathaccentV\macc@nested@a
  \if#31
    \macc@nested@a\relax111{#1}%
  \else
    \def\gobble@till@marker##1\endmarker{}%
    \futurelet\first@char\gobble@till@marker#1\endmarker
    \ifcat\noexpand\first@char A\else
      \def\first@char{}%
    \fi
    \macc@nested@a\relax111{\first@char}%
  \fi
  \endgroup
}
\makeatother

\begin{document}
\newcommand{\jw}[1]{{\textcolor{red}{[JW: #1]}}}
\title{Countering Mainstream Bias via End-to-End Adaptive Local Learning}
%
%


\author{Jinhao Pan\inst{1}\orcidID{0009-0006-1574-6376} \and
Ziwei Zhu\inst{2}\orcidID{0000-0002-3990-4774} \and \\
Jianling Wang\inst{1}\orcidID{0000-0001-9916-0976} \and Allen Lin\inst{1}\orcidID{0000-0003-0980-4323} \and \\James Caverlee\inst{1}\orcidID{0000-0001-8350-8528} }
\authorrunning{J. Pan et al.}
%



\institute{Texas A\&M University, College Station, TX, USA \and
George Mason University, Fairfax, VA, USA}
\maketitle              
%

\begin{abstract}
Collaborative filtering (CF) based recommendations suffer from mainstream bias -- where mainstream users are favored over niche users, leading to poor recommendation quality for many long-tail users. In this paper, we identify two root causes of this mainstream bias: (i) discrepancy modeling, whereby CF algorithms focus on modeling mainstream users while neglecting niche users with unique preferences; and (ii) unsynchronized learning, where niche users require more training epochs than mainstream users to reach peak performance. Targeting these causes, we propose a novel end-To-end Adaptive Local Learning (TALL) framework to provide high-quality recommendations to both mainstream and niche users. TALL uses a loss-driven Mixture-of-Experts module to adaptively ensemble experts to provide customized local models for different users. Further, it contains an adaptive weight module to synchronize the learning paces of different users by dynamically adjusting weights in the loss. Extensive experiments demonstrate the state-of-the-art performance of the proposed model. Code and data are provided at \url{https://github.com/JP-25/end-To-end-Adaptive-Local-Leanring-TALL-}


\end{abstract}
\section{Introduction}
\label{sec:intro}



The detrimental effects of algorithmic bias in collaborative filtering (CF) recommendations have been widely acknowledged~\cite{chen2023improving,chen2023fairly,wang2023survey,zhu2022fighting}. Among these different types of recommendation bias, an especially critical one is \textbf{Mainstream Bias} (also called ``grey-sheep'' problem)~\cite{alabdulrahman2021catering,gras2017can,li2021leave,zhu2022fighting}, which refers to the phenomenon that \textit{a CF-based algorithm delivers recommendations of higher utility to users with mainstream interests at the cost of poor recommendation performance for users with niche or minority interests}. For example, in a social media platform, mainstream users with interests in prevalent social topics will receive recommendations of high accuracy, while the system struggles to provide precise recommendations for niche users who focus on less common, yet equally important, topics. This makes the platform deliver unfair services to users with distinct interests and ultimately deteriorates the long-term prosperity of the platform. 


In this work, we identify two core root causes of such a mainstream bias: the discrepancy modeling problem and the unsynchronized learning problem.

\noindent\textbf{Discrepancy Modeling Problem:} CF algorithms estimate user preferences based on other users with similar tastes. So, the data from users with different preferences cannot help (or even play a negative role) in predicting recommendations for a target user. This issue functions bidirectionally -- it impacts niche users, who differ from the majority, and mainstream users affected by the data from niche users. While prior studies~\cite{choi2021local,zhu2022fighting} have used heuristic-based local learning to craft customized models for different user types (e.g., mainstream vs. niche), their efficacy is often bound by the quality of the underlying heuristics. This underscores the need for \textit{adaptive approaches that learn to generate locally customized models for different user types in an end-to-end fashion.}

\begin{wrapfigure}{r}{0.4\textwidth}
\vspace{-25pt}
\centering
\includegraphics[ width=1\linewidth ]{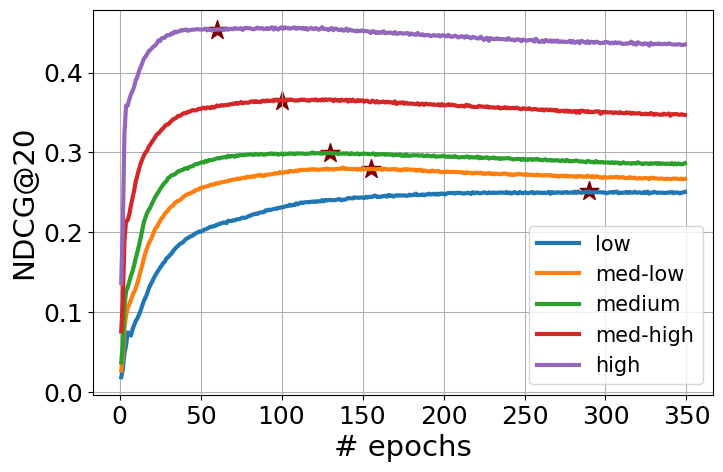} 
\vspace{-18pt} 
\caption{\small Validation \textit{NDCG@20} during training for users of varying mainstream levels. The star marks the epoch when reaching the peak performance of the subgroup.}
\label{fig:user_change_motif} 
\vspace{-22pt} 
\end{wrapfigure}

\noindent\textbf{Unsynchronized Learning Problem:} Another factor contributing to mainstream bias is the different learning paces between mainstream and niche users. Intuitively, mainstream users, with abundant training signals, tend to reach optimal learning faster. For instance, in Figure~\ref{fig:user_change_motif}, the high-mainstream subgroup peaks at around 60 epochs, whereas the low-mainstream group takes nearly 300 epochs. Training a model for all users without considering these learning pace disparities often results in models that cater primarily to mainstream users, sidelining niche users from reaching their optimal utility. Addressing this requires a method to \textit{synchronize the learning process across users, irrespective of their mainstreamness}.

To address these problems, we propose an end-\underline{\textbf{T}}o-end \underline{\textbf{A}}daptive \underline{\textbf{L}}ocal \underline{\textbf{L}}earning (TALL) framework. To tackle the discrepancy modeling problem affecting both niche and mainstream users, we devise a loss-driven Mixture-of-Experts structure as the backbone. This structure achieves local learning via an end-to-end neural network and adaptively assembles expert models using a loss-driven gate model, offering tailored local models for different users. Further, we develop an adaptive weight module to dynamically adjust the learning paces by weights in the loss, ensuring optimal learning for all user types. With these two complementary and adaptive modules, TALL can effectively \textbf{promote the utility for niche users while preserving or even elevating the utility for mainstream users}, leading to a significant debiasing effect based on the \textbf{Rawlsian Max-Min fairness principle}~\cite{rawls2001justice}.

In sum, our contributions are: (1) we propose a loss-driven Mixture-of-Experts structure to tackle the discrepancy modeling problem, highlighted by an adaptive loss-driven gate module for customized local models; (2) we introduce an adaptive weight module to synchronize learning paces, augmented by a loss change and a gap mechanism for better debiasing; and (3) Extensive experiments demonstrate TALL’s superior debiasing capabilities compared to leading alternatives, enhancing utility for niche users by 6.1\% over the best baseline with equal model complexity. Data and code are available at \url{https://github.com/JP-25/end-To-end-Adaptive-Local-Leanring-TALL-}.



\section{Preliminaries}

\noindent\textbf{Problem Formalization.} Given a user set $\mathcal{U}=\{1,2,\ldots, N\}$ consisting of $N$ users and an item set $\mathcal{I}=\{1,2,\ldots, M\}$ consisting of $M$ items, we have the implicit feedback from users to items as the set $\mathcal{O}=\{(u,i)\}$, where $u\in\mathcal{U}$ refers to a user and $i\in\mathcal{I}$ refers to an item. This feedback set serves as the training data for training a recommendation model that generates recommendations for users. Each user $u$ is represented by a binary vector of length $M$, denoted as $\mathbf{O}_u\in\{0,1\}^M$. During the evaluation, the model provides a ranked list of $K$ recommended items for every user. Various ranking evaluation metrics can be employed, such as NDCG@$K$ and Recall@$K$~\cite{liang2018variational}.

\noindent\textbf{Mainstream Bias.} Typically, a recommender system is evaluated by averaging the utility over all users (such as NDCG@$K$), which essentially conceals the performance differences across different types of users. Previous work~\cite{zhu2022fighting} formalizes the mainstream bias as \textit{the recommendation performance difference across users of different levels of mainstreamness}. In this work, we follow the problem setting of~\cite{zhu2022fighting} to measure the mainstream level for a user by calculating the average similarity of the user to all others: the more similar the user is to the majority, the more mainstream she is.

\noindent\textbf{Debiasing Goal.} In terms of the goal of debiasing, while addressing the issues of discrepancy modeling and unsynchronized learning benefits all user types, it is inappropriate to expect equalized utility across all users, which possibly encourages decreasing the utility for mainstream groups. Hence, in this work, we follow the \textbf{Rawlsian Max-Min fairness} principle of distribute justice~\cite{rawls2001justice}. To achieve fairness, this principle aims to maximize the minimum utility of individuals or groups, ensuring that no one is underserved. So, to counter the mainstream bias, we aim to \textbf{improve the recommendation utility of niche users while preserving or even enhancing the performance of mainstream users}.

\section{End-to-End Adaptive Local Learning}
\label{sec:debias} 

\vspace{-5pt}

\begin{figure}[t!]
\centering
\includegraphics[width=1\linewidth ]{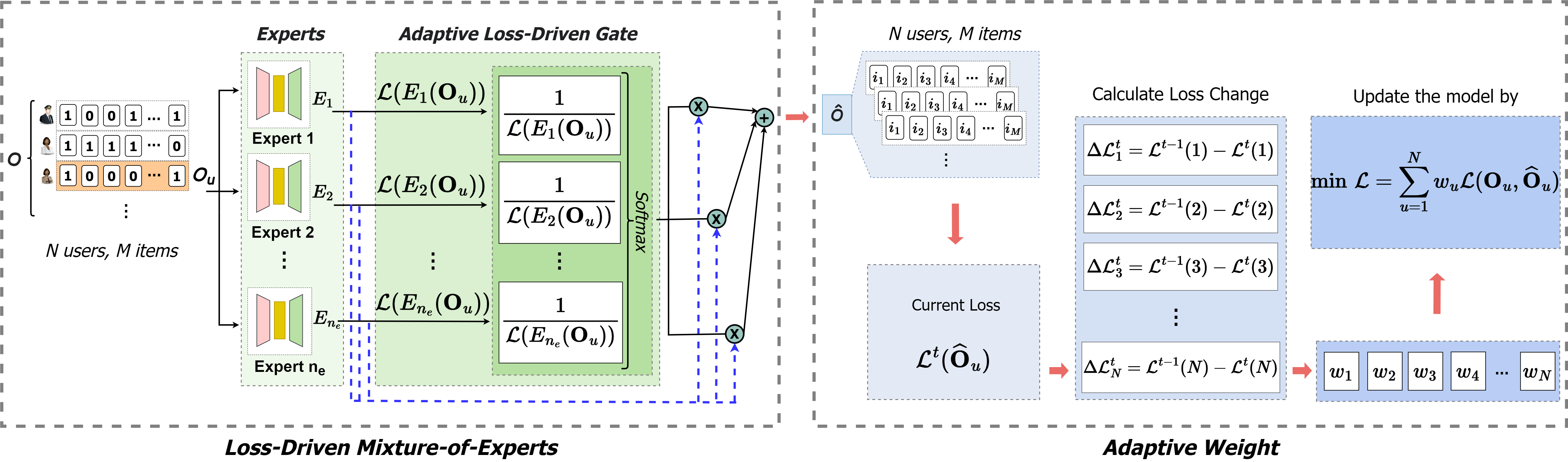} 
\vspace{-15pt} 
\caption{The proposed End-to-End Adaptive Local Learning (TALL) framework.}
\label{fig:TALL} 
\vspace{-15pt}
\end{figure}

To debias, we propose the end-\textbf{T}o-end \textbf{A}daptive \textbf{L}ocal \textbf{L}earning (\textbf{TALL}) framework, shown in Figure~\ref{fig:TALL}. To address the \textit{discrepancy modeling} problem, this framework integrates a \textbf{loss-driven Mixture-of-Experts} module to adaptively provide customized models for different users by an end-to-end learning procedure. To address the \textit{unsynchronized learning} problem, the framework involves an \textbf{adaptive weight} module to synchronize the learning paces of different users by adaptively adjusting weights in the objective function.

\subsection{Loss-Driven Mixture-of-Experts}
\label{sec:L-MoE}

\noindent\textbf{Mixture-of-Experts.}
\label{sec:MoE}
To address the discrepancy modeling problem, prior works propose the local learning method~\cite{choi2021local,zhu2022fighting} to provide a customized local model trained by a small collection of local data for each user. However, how to curate such a local dataset and how to build a local model are completely hand-crafted in these algorithms, which significantly limits performance.  
To overcome this weakness, we adopt a Mixture-of-Experts (MoE)~\cite{eigen2013learning} structure as the backbone of our proposed framework to implement end-to-end local learning.

In detail, the MoE has two main components: gate and expert models. Specifically, an MoE comprises multiple expert models, each of which is trained to work for a specific task. For the recommendation problem, we adopt the MultVAE~\cite{liang2018variational} as the expert model, and each MultVAE-based expert model in the MoE is responsible to process certain types of users. 
Then, with these expert models, for a target user, we rely on the gate model to distribute gate values to different expert models and generate the prediction for the user by weighted averaging the outputs of expert models based on gate values. 
Concretely, given an input user $u$ with feedback record $\mathbf{O}_{u}$, the output of the MoE is $\widehat{\mathbf{O}}_{u}=\sum_{k=1}^{n_e} G_k(\mathbf{O}_{u})E_k(\mathbf{O}_{u})$, 
where $n_e$ is the number of expert models, $G_k(\mathbf{O}_{u})$ and $E_k(\mathbf{O}_{u})$ are the $k$-th value of the gate model output and the output of the $k$-th expert model, and we also have the constraint $\sum_{k=1}^{n_e} G_k(\mathbf{O}_{u}) = 1$. The expert and gate models are trained together by the data in an end-to-end fashion.

\noindent\textbf{Adaptive Loss-Driven Gate.} 
\label{sec:L-gate}
However, the regular gate model, a free-to-learn feed-forward neural network (i.e., a multilayer perception) within a standard MoE, is also susceptible to various biases, including mainstream bias. The gate model is trained by data with more mainstream users and thus focuses more on how to assign gate values to improve utility for mainstream users while overlooking niche users. This results in an inability of the regular gate model to reasonably assign values to different users, especially niche users. Therefore, such a free-to-learn gate model cannot address the discrepancy modeling problem. A more precise and unbiased gate mechanism is needed.

A key principle of the gate model is that when we have a set of expert models, the gate model should assign high values for expert models that are effective for the target user, and low values for expert models less effective for the target user. In this regard, the loss function serves as a high-quality indicator. Specifically, a high loss value in an expert model for a target user means that this expert model is not helpful for delivering prediction, and thus, it should not contribute to the aggregation of the final customized local model, receiving a small gate value. Conversely, a low loss indicates an effective expert model deserving a higher gate value. Based on this intuition, we propose the adaptive loss-driven gate module:
\begin{equation*}\small
\begin{aligned}
\centering
G_k(\mathbf{O}_u)={e^{(\mathcal{L}(E_k(\mathbf{O}_u)))^{-1}}} \Big/ {\sum_{t=1}^{n_e}e^{(\mathcal{L}(E_t(\mathbf{O}_u)))^{-1}}},
\end{aligned}
\label{equ:gate_eq}
\end{equation*}
where $\mathcal{L}(E_k(\mathbf{O}_u)))$ is the loss function of the $k$-th expert model for user $u$, and we adopt the multinomial cross-entropy loss function from \cite{liang2018variational}.


This mechanism complies with the principle of allocating high gate values to more effective expert models and low values to irrelevant expert models. In practice, we can use the loss function calculated on training data or independent validation data for the proposed adaptive loss-driven gate mechanism. Due to that loss on previously unseen validation data is a more precise signal of model performance, in this work, we adopt the loss function on validation data to calculate gate values. By this adaptive loss-driven gate mechanism, we can automatically and adaptively assign gate values to expert models, based on which we generate effective customized models.

\subsection{Synchronized Learning via Adaptive Weight}
\label{sec:synchronized_learning}

After addressing the discrepancy modeling problem, another core cause of the mainstream bias is the unsynchronized learning problem -- the learning difficulties vary for different users, and users reach the performance peak during training at different speeds (refer to Figure~\ref{fig:user_change_motif} for an example). Thus, a method to synchronize the learning paces of users is desired. We devise an adaptive weight approach to achieve learning synchronization so that the dilemma of performance trade-off between mainstream users and niche users can be overcome.


\noindent\textbf{Adaptive Weight.}
\label{sec:AW}
The fundamental motivation of the proposed method lies in linking the learning status of a user to the loss function of the user at the current epoch. A high loss for a user signifies ineffective learning by the model, necessitating more epochs for accurate predictions. Conversely, a low loss indicates successful modeling, requiring less or even no further training. Moreover, we can use a weight in the loss function to control the learning pace for the user: a small weight induces slow updating, and a large weight incurs fast updating. 
Hence, we propose to synchronize different users by applying weights to the objective function based on losses users get currently -- a user with a high loss should receive a large weight, and vice versa. We aim to achieve this intuition by solving the following optimization problem:
\begin{equation}\small
\begin{aligned}
\max_{w}\sum_{u=1}^{N}w_{u}\mathcal{L}(\mathbf{O}_{u}, \widehat{\mathbf{O}}_{u})-\alpha\|w\|_{2}^{2},\,\,\,\,\,\,\,\,\,\,\,  \text{s.t.}\,\,\,& \sum_{u=1}^{N}w_{u}=N, w \geq 0,
\end{aligned}
\label{equ:adaptive_weight_optimization}
\end{equation}
where we aim to maximize the weighted sum of losses of users, in which the solution will assign high weights to users with large losses; and the regularization $\alpha\|w\|_{2}^{2}$ is to control the skewness of the weight distribution -- a larger $\alpha$ leads to a more even distribution. One extreme case is when $\alpha\rightarrow+\infty$, $w_u=1$ for all users, and the other extreme case is when $\alpha=0$, $w_u=N$ for user $u$ with the largest loss. To solve this optimization problem, a closed-form solution by Lagrange Multipliers with Karush–Kuhn–Tucker (KKT) conditions can be derived~\cite{chai2022fairness}:
\begin{equation}\small
\begin{aligned}
\centering
w_{u}^{*}=\max \left( \left (\mathcal{L}(\mathbf{O}_{u}, \widehat{\mathbf{O}}_{u})-\lambda \right ) \Big/ 2 \alpha, 0\right),\,\,\,\,\,\,\,\,\,\,\lambda=\left(\sum_{v=1}^{N} \mathcal{L}(\mathbf{O}_{v}, \widehat{\mathbf{O}}_{v})-2\alpha N \right ) \Big/ N.
\end{aligned}
\label{equ:original_aw}
\end{equation}

The proof for this solution can be found in~\cite{chai2022fairness}. Finally, with the computed weights, we can insert them into the original learning objective function of the framework and train the model by minimizing $\sum_{u=1}^N w_u\mathcal{L}(\mathbf{O}_{u}, \widehat{\mathbf{O}}_{u})$. Similar to the proposed adaptive loss-driven gate model in Section~\ref{sec:L-gate}, here we also rely on losses calculated by validation data to compute high-quality weights.

\noindent\textbf{Loss Change Mechanism and Gap Mechanism.}
With the proposed adaptive weight approach, we take a solid step toward learning synchronization. However, two critical issues remain unaddressed. First, the scale of the loss function is innately different across different users. Usually, mainstream users possess a lower loss value than niche users because the algorithm can achieve better utility for mainstream users who are easier to model. Due to this \textit{scale diversity problem}, computing weights by exact values of the loss can lead to the undesired situation that mainstream users get overly low weights and niche users get overly high weights, disturbing the learning process. On the other hand, the loss is not always stable, especially at the early stage of training. This \textit{unstable loss problem} can deteriorate the efficacy of the problem adaptive weight method too.  

To tackle the scale diversity problem, instead of directly using the loss, we propose to use loss change across epochs as the indicator for computing weights. If we observe the change of loss in recent epochs for a user is significant and the loss is in a decreasing trend, then we conclude that this user is in a fast learning stage and needs more epochs to converge. And we want to assign a high weight to this user. Thus, after each epoch, we will record the loss change $\Delta \mathcal{L}^{t}_u = \mathcal{L}^{t-1}(u) - \mathcal{L}^{t}(u)$. Furthermore, to make the indicator more robust to counter the scale diversity problem, in our experiment, we average recent $L$ loss change values, denoted as $\widebar{\Delta\mathcal{L}}^{t}_u$, to replace the loss in Equation~\ref{equ:adaptive_weight_optimization}. We can then derive the solution:
\begin{equation} \small
\begin{aligned}
\centering
w_{u}^{*}=\max\left( \left (\widebar{\Delta\mathcal{L}}^{t}_u-\lambda \right ) \Big/ 2 \alpha, 0\right),\,\,\, \lambda=\left (\sum_{v=1}^{N} \widebar{\Delta\mathcal{L}}^{t}_v-2\alpha N \right ) \Big/ N.
\end{aligned}
\label{equ:aw_solution}
\end{equation}

At last, since the losses are excessively unstable at the initial stage of training and the proposed adaptive weight module heavily relies on the stability of the loss value, the proposed method cannot perform well at the initial stage of training. Hence, we propose to have a gap at the beginning for our adaptive weight method. That is, we do not apply the proposed adaptive weight method to the framework at the first $T$ epochs. Since at the early stage of the training, all users will be at a fast learning status (consider the first 50 epochs in Figure~\ref{fig:user_change_motif}), as there is no demand for learning synchronization. After $T$ epochs of ordinary training, when the learning procedure is more stable and the loss is more reliable, we apply the adaptive weight method to synchronize the learning for different users. And this time synchronization is desired and plays an important role. The gap window $T$ is a hyper-parameter and needs to be predefined.

\section{Debiasing Experiments}
\label{sec:Experiments}
\begin{wraptable}{r}{5.2cm}\footnotesize 
\vspace{-30pt}
\caption {Data statistics.}
\vspace{-10pt}
\centering
\begin{adjustbox}{width=0.4\textwidth}
\begin{tabular}{|c|ccc|}
\hline
        & \#users & \#items & density \\ \hline
ML1M    & 6,040   & 3,706   & 4.46\%  \\
Yelp    & 20,001  & 7,643   & 0.32\%  \\
CDs \& Vinyl & 12,023  & 8,050   & 0.32\%  \\ \hline
\end{tabular}
\end{adjustbox} 
\label{table:datasets}
\vspace{-22pt}
\end{wraptable}

In this section, we present a comprehensive set of experiments to highlight the strong debiasing performance of the proposed method, validate the effectiveness of various model components, assess the impact of the proposed adaptive weight module, and examine the impact of hyper-parameters.

\subsection{Experimental Setup}

\noindent\textbf{Data and Metric.} Table~\ref{table:datasets} summarizes statistics of datasets used in this paper. We use three public datasets for the experiments: \textbf{ML1M}~\cite{harper2015movielens}, \textbf{Yelp}~\cite{yelp}, and \textbf{Amazon CDs and Vinyl}~\cite{ni2019justifying}. For each dataset, we consider the ratings or reviews as positive feedback from users to items. Then, following the same evaluation scheme from \cite{zhu2022fighting}, for each dataset, we uniformly randomly divide it into training, validation, and testing sets in the ratio of 70\%, 10\%, and 20\%. Next, we calculate the mainstream scores of users. Specifically, for a user $u$, the mainstream score is calculated as $MS_u=\sum_{v\in\mathcal{U}\setminus u}Sim(\mathbf{O}_u, \mathbf{O}_v)/(N-1)$, 
where $Sim(\mathbf{O}_u, \mathbf{O}_v)$ is the user-user similarity between users $u$ and $v$. The similarity is computed by Jaccard similarity between the implicit feedback record $\mathbf{O}_u$ and $\mathbf{O}_v$. Then, we sort users based on calculated mainstream scores in non-descending order and divide them into five subgroups with equal sizes. 
In the result, we denote the first 20\% of users with the lowest mainstream scores as users of `low' mainstream level, the subgroup of 20\%-40\% users as `med-low' mainstream level, and so on for 40\%-60\% (`medium'), 60\%-80\% (`med-high'), and 80\%-100\% (`high') users. Last, we report and compare the average NDCG@20 for each subgroup to show the mainstream bias. We do not divide users by cutting at specific mainstream scores because this could lead to groups with extremely small numbers of users, deteriorating the reliability of reported utility and bias evaluation.

Considering the Rawlsian Max-Min fairness principle~\cite{rawls2001justice}, \textbf{the goal of debiasing is to promote the average NDCG@20 for subgroups with low mainstream scores while preserving or even improving the utility for subgroups with high mainstream scores at the same time. Hence, we also anticipate an increase in the overall NDCG@20 of the model.}

\smallskip
\noindent\textbf{Baselines.}
In the experiments, we compare the proposed TALL with MultVAE and four state-of-art debiasing methods: (1) \textbf{MultVAE}~\cite{liang2018variational} is the widely used vanilla recommendation model without debiasing. (2) \textbf{WL}~\cite{zhu2022fighting} is a global method designed to assign more weights to niche users in the training loss of MultVAE. (3) \textbf{LOCA}~\cite{choi2021local} is a local learning model that trains multiple anchor models corresponding to identified anchor users and aggregates the outputs from anchor models based on the similarity between the target user and anchor users. (4) \textbf{LFT}~\cite{zhu2022fighting} is the SOTA local learning model that first trains a global model with all data and then fine-tunes a customized local model for each user using their local data for the target user. (5) \textbf{EnLFT}~\cite{zhu2022fighting} is the ensembled version of LFT, which is similar to LOCA but trains the anchor models by the approach of LFT.

To fairly compare the performance of different baselines and our proposed model, for all four debiasing baselines and our proposed TALL, we adopt the MultVAE as the base model (or the expert model in TALL). LOCA, EnLFT, and TALL have the same complexity with a fixed number of MultVAE in them. And owing to the end-to-end training paradigm, TALL takes less training time than other local learning baselines. Last, LFT has the largest complexity, which has an independent MultVAE for each user.

\begin{table*}[t!]
\caption{\small Comparing TALL with SOTA debiasing baselines on 3 datasets.}
\centering
\begin{adjustbox}{width=\textwidth}

\begin{tabular}{|c|c|ccccc|c|ccccc|c|ccccc|}
\hline
\multirow{2}{*}{} & \multicolumn{6}{c|}{ML1M} & \multicolumn{6}{c|}{Yelp} & \multicolumn{6}{c|}{CDs \& Vinyl} \\ \cline{2-19}
& \multirow{2}{*}{\makecell{NDCG\\@20}} & \multicolumn{5}{c|}{Subgroups of mainstream levels} & \multirow{2}{*}{\makecell{NDCG\\@20}} & \multicolumn{5}{c|}{Subgroups of mainstream levels} & \multirow{2}{*}{\makecell{NDCG\\@20}} & \multicolumn{5}{c|}{Subgroups of mainstream levels} \\
& & L & ML & M & MH & H & & L & ML & M & MH & H & & L & ML & M & MH & H \\ \hline
MultVAE & .3260 & .2354 & .2764 & .2986 & .3652 & .4546 & .0877 & .0686 & .0710 & .0733 & .0901 & .1355 & .1367 & .1100 & .1316 & .1366 & .1457 & .1596 \\
WL & .3278 & .2448 & .2801 & .2970 & .3639 & .4532 & .0870 & .0700 & .0708 & .0720 & .0888 & .1332 & .1361 & .1133 & .1334 & .1361 & .1441 & .1534 \\
EnLFT & .3341 & .2586 & .2875 & .3025 & .3661 & .4556 & .0887 & .0697 & .0715 & .0740 & .0915 & .1369 & .1453 & .1230 & .1387 & .1423 & .1561 & .1666 \\
LOCA & .3308 & .2551 & .2780 & .2972 & .3622 & .4617 & .0942 & .0723 & .0758 & .0764 & .0970 & .1494 & .1573 & .1341 & .1510 & .1556 & .1665 & .1795 \\
LFT & .3416 & .2707 & \textbf{.2918} & .3072 & .3727 & .4657 & .0927 & .0740 & .0738 & .0768 & .0956 & .1432 & .1557 & .1343 & .1481 & .1515 & .1678 & .1770 \\
TALL & \textbf{.3456} & \textbf{.2746} & .2903 & \textbf{.3112} & \textbf{.3784} & \textbf{.4734} & \textbf{.0992} & \textbf{.0772} & \textbf{.0803} & \textbf{.0826} & \textbf{.1056} & \textbf{.1505} & \textbf{.1700} & \textbf{.1392} & \textbf{.1599} & \textbf{.1652} & \textbf{.1844} & \textbf{.2013} \\ \hline\hline
$\Delta_{MultVAE}(\%)$ & 6.01 & 16.65 & 5.03 & 4.22 & 3.61 & 4.14 & 13.11 & 12.54 & 13.1 & 12.69 & 17.2 & 11.07 & 24.36 & 26.54 & 21.5 & 20.94 & 26.56 & 26.13 \\ \hline
$\Delta_{LFT}(\%)$ & 1.17 & 1.44 & -0.51 & 1.30 & 1.53 & 1.65 & 7.01 & 4.32 & 8.81 & 7.55 & 10.46 & 5.10 & 9.18 & 3.65 & 7.97 & 9.04 & 9.89 & 13.73 \\ \hline
$\Delta_{LOCA}(\%)$ & 4.47 & 7.64 & 4.42 & 4.71 & 4.47 & 2.53 & 5.31 & 6.78 & 5.94 & 8.12 & 8.87 & 0.74 & 8.07 & 3.80 & 5.89 & 6.17 & 10.75 & 12.14 \\ \hline
\end{tabular}

\end{adjustbox}

{\small L: low, ML: med-low, M: medium, MH: med-high, H: high}
\label{table:all}
\vspace{-20pt}
\end{table*}

\smallskip
\noindent\textbf{Reproducibility.} All models are implemented in PyTorch and optimized by the Adam algorithm~\cite{kingma2014adam}. For the baseline MultVAE and the MultVAE component in other models, we set one hidden layer of size 100. And we maintain the number of local models at 100 for LOCA, EnLFT, and TALL for all datasets to ensure a fair comparison. All other hyper-parameters are grid searched by the validation sets. \textit{All code and data can be found at \url{https://github.com/JP-25/end-To-end-Adaptive-Local-Leanring-TALL-}}.


\subsection{Debiasing Performance}
\label{sec:experiment_debias}
\vspace{-5pt}

First, we conduct a comparative analysis to show the effectiveness of the proposed TALL. In Table~\ref{table:all}, we evaluate the overall NDCG@20 and average NDCG@20 for five user subgroups with varying mainstream levels for all methods and datasets. The best results of each metric and subgroup for all datasets are marked in bold, and the improvement rate of the proposed TALL over the best baseline MultVAE, LOCA, and LFT is exhibited as well. The user subgroups are categorized based on their mainstream scores.




\noindent\textbf{TALL vs. MultVAE \& WL.} First, we can observe that the utilities for users of all five subgroups are greatly promoted by our proposed TALL compared to the widely used model MultVAE. Moreover, we can see that although the global debiasing method WL can alleviate the mainstream bias to a certain degree compared to MultVAE, our proposed TALL can produce higher NDCG@20 for all five user subgroups than WL, depicting that the proposed TALL exhibits a more outstanding debiasing ability over WL. 


\noindent\textbf{TALL vs. EnLFT \& LOCA.} Hence, we next have a fair comparison between models of the same complexity and compare directly across local learning methods. From Table~\ref{table:all}, we observe that LOCA and EnLFT are more effective in mitigating the mainstream bias than WL, as they remarkably enhance utility across all five groups on all datasets. Meanwhile, our TALL significantly outperforms LOCA and EnLFT across all user groups and datasets. The improvement is especially prominent for niche users: TALL improves NDCG@20 of the `low' user group by 6.07\% on average over LOCA and 10\% over EnLFT. This shows that with the same model complexity, the proposed end-to-end adaptive local learning model is more effective than heuristic-based local learning models.


\noindent\textbf{TALL vs. LFT.} Last, we compare TALL with the state-of-the-art local learning baseline LFT, which 
is heavily computationally intense and time-consuming. But due to its special design that every user gets their own customized model trained by their local data, LFT can effectively address the discrepancy modeling problem. From Table~\ref{table:all}, we observe that, for most of the time, LFT achieves the best performance for niche users among all baselines. And LFT can perform especially effectively for dense datasets (i.e., ML1M). In fact, given that the model complexity of LFT is much higher than TALL, it is unfair to compare them only based on recommendation accuracy (i.e., NDCG). For example, in the ML1M dataset, our TALL contains 100 expert models (MultVAE), while LFT trains 6,040 (\#users) models separately, which is over 60 times larger than TALL. Although it is not a fair comparison, we can still observe in Table~\ref{table:all} that TALL can outperform LFT in most cases. Especially for the two sparse datasets Yelp and CDs\&Vinyl, TALL produces significantly higher utilities for all types of users. This demonstrates the efficacy and necessity of an end-to-end local learning method compared to a heuristic-based one.

In sum, from Table~\ref{table:all}, we see that for all datasets, the proposed TALL produces the greatest NDCG@20 improvement for each subgroup of different mainstream levels and leads to the state-of-the-art overall model performance. TALL can outperform baselines with lower and the same model complexity, and it can even outperform the baseline model that is way more complex than it.

\subsection{Ablation Study}
Next, we aim to investigate the effectiveness of different components in the proposed framework, including the proposed adaptive loss-driven gate module, the adaptive weight module, the gap mechanism in the adaptive weight module, and the loss change mechanism in the adaptive weight module.


\begin{table}[t!]\footnotesize
\caption {Ablation study on the adaptive loss-driven gate.}
\centering
\begin{adjustbox}{width=0.6\textwidth}

\begin{tabular}{|cc|c|ccccc|}
\hline
                                              &         &                 & \multicolumn{5}{c|}{Subgroups of different mainstream levels}        \\
                                              &         & NDCG@20         & L          & ML         & M         & MH          & H        \\ \hline
\multicolumn{1}{|c}{}    & MultVAE  & 0.3260    & 0.2354    & 0.2764          & 0.2986          & 0.3652    & 0.4546    \\
\multicolumn{1}{|c}{}                         & MoE & 0.3230          & 0.2513          & 0.2668    & 0.2895    & 0.3519          & 0.4553          \\
\multicolumn{1}{|c}{}                         & LMoE      & \textbf{0.3401} & \textbf{0.2714} & \textbf{0.2893} & \textbf{0.3066} & \textbf{0.3736} & \textbf{0.4594} \\ \hline
\end{tabular}
\end{adjustbox}

{\small L: low, ML: med-low, M: medium, MH: med-high, H: high}
\label{table:loss_driven_moe}
\vspace{-10pt}
\end{table}

\noindent \textbf{Adaptive Loss-Driven Gate.} To verify the effectiveness of the proposed adaptive loss-driven gate module in the proposed TALL, we compare the MoE component with the adaptive loss-driven gate (denoted as LMoE, which is the TALL model without the adaptive weight module) to a conventional MoE (denoted as MoE) with the standard multilayer perceptron (MLP) as the gate learning from the dataset. By comparing LMoE and MoE, we can justify the effect of the proposed adaptive loss-driven gate. The results are present in Table~\ref{table:loss_driven_moe}, where we also include the result of MultVAE as a baseline. From the table, we can observe that MoE produces better performance for niche users compared to MultVAE but worse results for other users. This is caused by the MLP-based gate model in MoE, which cannot precisely allocate gate values to expert models to ensemble a strong customized model for different users. Conversely, we can see that even without the adaptive weight module, the LMoE can deliver greatly higher utilities for all types of users compared to MultVAE and MoE, showing the strong capability of the proposed adaptive loss-driven gate module in distributing gate values across expert models. And this result also demonstrates the efficacy of the proposed LMoE module in TALL in terms of addressing the discrepancy modeling problem.

\begin{table}[t!]\footnotesize
\caption {Ablation study on the adaptive weight module, gap mechanism, and loss change mechanism.}
\centering
  \begin{adjustbox}{width=0.77\textwidth}

\begin{tabular}{|cc|c|ccccc|}
\hline
                                              &         &                 & \multicolumn{5}{c|}{Subgroups of different mainstream levels}        \\
                                              &         & NDCG@20         & L          & ML         & M         & MH          & H       \\ \hline
\multicolumn{1}{|c}{}                            & LMoE  & 0.3401    & 0.2714    & 0.2893          & 0.3066          & 0.3736    & 0.4594    \\
\multicolumn{1}{|c}{}                         & LMoE + LC   & 0.3371          & 0.2566          & 0.2835    & 0.3075    & 0.3724          & 0.4654          \\
\multicolumn{1}{|c}{}                         & LMoE + gap + L & 0.3396          & 0.2681          & 0.2878    & 0.3065    & 0.3741          & 0.4616          \\
\multicolumn{1}{|c}{}                         & TALL (LMoE + gap + LC)      & \textbf{0.3456} & \textbf{0.2746} & \textbf{0.2903} & \textbf{0.3112} & \textbf{0.3784} & \textbf{0.4734} \\ \hline
\end{tabular}
\end{adjustbox}

{\small L: low, ML: med-low, M: medium, MH: med-high, H: high}
\label{table:adaptive_weights}
\vspace{-20pt}
\end{table}

\noindent \textbf{Adaptive Weight.} To address the unsynchronized learning problem, we develop the adaptive weight module to dynamically adjust the learning paces of different users. To verify the effectiveness of the proposed adaptive weight, we compare the complete TALL algorithm (with both the loss-driven MoE module and the adaptive weight module) to the loss-driven MoE module (LMoE). The results on ML1M are shown in Table~\ref{table:adaptive_weights}, from which we can see that TALL outperforms LMoE for all types of users, manifesting the effectiveness of the proposed adaptive weight method. Furthermore, we explore the effectiveness of two special mechanisms, the gap mechanism and the loss change mechanism within the adaptive weight module, in the subsequent sections.

\noindent \textbf{Gap Mechanism.} To avoid the unstable loss problem at the initial training stage, we propose to have a gap for the adaptive weight method, i.e., we wait for a certain number of epochs at the initial training stage until the loss is stable and then apply the adaptive weight method. To verify the effectiveness of such a gap strategy, we compare the complete TALL (with a full version of the adaptive weight module including both the gap mechanism and the loss change mechanism) to a variation of TALL with the adaptive weight module without the gap mechanism. The comparison is presented in Table~\ref{table:adaptive_weights} as LMoE+LC vs. TALL. We can observe that the gap mechanism does have a significant influence on the model performance that the model with the gap mechanism (TALL) delivers better utilities for all types of users than the model without the gap mechanism (LMoE+LC).


\noindent \textbf{Loss Change Mechanism.} Last, we aim to verify the effectiveness of the proposed loss change mechanism in the adaptive weight module. The goal of loss change is to counter the scale diversity problem when applying the adaptive weight module. Here, we compare the complete TALL (with a full version of the adaptive weight module including both the gap mechanism and the loss change mechanism) to a variation of TALL (LMoE+gap+L) with the adaptive weight module that uses original loss as introduced in Section~\ref{sec:AW}. From the comparison result shown in Table~\ref{table:adaptive_weights}, we see that the model using loss change (TALL) performs better than the model using original loss for calculating weights (LMoE+gap+L). TALL outperforms LMoE+gap+L for all types of users.


In sum, by a series of comparative analyses, we show that the proposed adaptive loss-driven gate module, adaptive weight module, the gap mechanism in adaptive weight, and the loss change mechanism in adaptive weight are effective and play imperative roles in the proposed TALL framework.

\begin{wrapfigure}{r}{0.5\textwidth}
    \vspace{-30pt}
    \centering
    \includegraphics[ width=1\linewidth ]{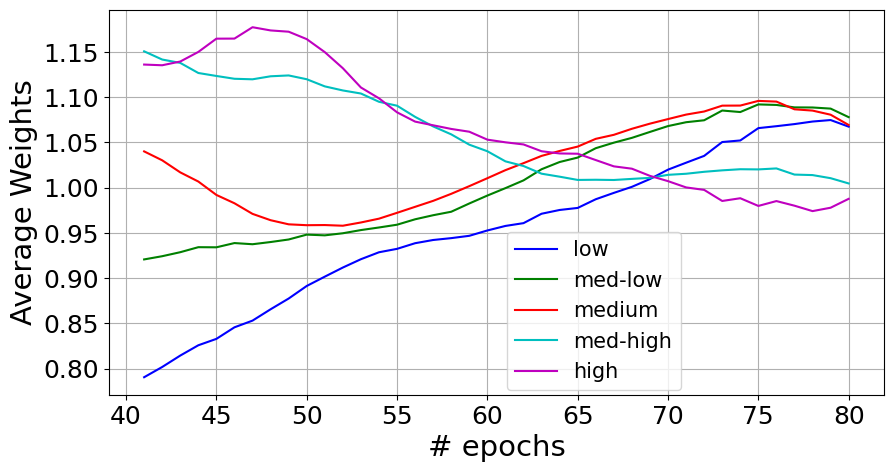} 
    \vspace{-20pt} 
    \caption{Weights assignment across different types of users.}
    \label{fig:avg_weights} 
    \vspace{-20pt} 
\end{wrapfigure}

\subsection{Effect of the Adaptive Weight Module}
Last, we turn our attention to investigating the effect of the adaptive weight module, studying how it synchronizes the learning paces of different users. We run TALL on the ML1M dataset and present the average weights for the five subgroups with the gap window ($\#gap=40$) in Figure~\ref{fig:avg_weights}. It can be observed that the adaptive weight module assigns weights dynamically to different types of users to synchronize their learning paces. Initially, mainstream users receive higher weights because they are easier to learn and have a higher upper bound of performance than niche users. Then, when mainstream users reach the peak, the model switches the attention to niche users who are more difficult to learn, gradually increasing the weights for `low', `med-low', and `medium' users until the end of the training procedure. However, `med-high' and `high' users, approaching converged, need a slower learning pace to avoid overfitting, leading to a decrease in the weights. 
Figure~\ref{fig:avg_weights} illuminates the effectiveness and dynamic nature of the proposed adaptive weight module in synchronizing the learning procedures for different types of users.

\subsection{Hyper-parameter Study}
\label{sec:Hyper-parameter_Study}

Additionally, we have also conducted a comprehensive hyper-parameter study investigating the impacts of three hyper-parameters in TALL: (1) the gap window in the adaptive weight module; (2) $\alpha$ in the adaptive weight module; and (3) the number of experts. The complete results are in \url{https://github.com/JP-25/end-To-end-Adaptive-Local-Leanring-TALL-/blob/main/Hyperparameter_Study.pdf}.

\section{Related Work}
\label{sec:related_work}
Fairness and bias issues in recommender systems have attracted increasing attention recently. Popularity bias~\cite{abdollahpouri2017controlling,abdollahpouri2019managing,chen2023graph,wei2020model,zhang2021causal,zhang2021model}, exposure bias~\cite{ben2023learning,lee2021dual,saito2020unbiased,saito2020unbias}, and item fairness~\cite{beutel2019fairness,cai2023causal,chen2023fairly,geyik2019fairness,liu2018personalizing,yao2017beyond} exemplify significant item-side biases. Besides prior works mainly focusing on the item perspective, several research studies have explored user biases, analyzing utility differences among diverse user groups based on user demographic attributes, like age or gender~\cite{chen2023improving,ekstrand2018all,fu2020fairness,li2021user,schedl2019online,wang2023survey,ying2023camus}. For instance, Ekstrand et al. \cite{ekstrand2018all} empirically investigated multiple recommendation models and demonstrated utility differences across user demographic groups. Schedl et al. \cite{schedl2019online} examined music preference differences among user age groups, revealing variations in recommendation performance. To address these issues, Fu et al. \cite{fu2020fairness} proposed to leverage rich information from knowledge graphs, Li et al. \cite{li2021user} developed a re-ranking to narrow the utility gap between different user groups, and Chen et al. \cite{chen2023improving} implemented data augmentation by generating ``fake'' data to achieve a balanced distribution.

However, demographic attributes may not comprehensively capture user interests and behaviors. Unlike the aforementioned works focusing on bias analysis based on demographic groups, \textbf{mainstream bias} poses a critical challenge in recommender systems.  Previous works~\cite{alabdulrahman2021catering,gras2017can,li2021leave} acknowledge mainstream bias as the ``grey-sheep'' problem, where ``grey-sheep users'' with niche interests lead to challenges in finding similar peers and result in poor recommendations. However, they do not propose robust bias measurements and debiasing methods. A more aligned study with better mainstream bias evaluations to this paper is~\cite{zhu2022fighting}, which also addresses mainstream bias and enhances utility for niche users using global and local methods. Prior existing local methods~\cite{choi2021local,christakopoulou2018local,lee2013local,lee2016llorma,zhu2022fighting} and global methods~\cite{zhu2022fighting} can mitigate the bias to some degree by improving the utility for niche users. The recently proposed Local Fine Tuning (LFT)~\cite{zhu2022fighting} and local collaborative autoencoder (LOCA)~\cite{choi2021local} produce state-of-the-art performance by employing multiple multinomial variational autoencoders (MultVAE)~\cite{liang2018variational} as base models and generating customized local models to capture special patterns of different types of user. Nonetheless, prior methods have a key limitation: their reliance on heuristics impacts performance, necessitating meticulous hyper-parameters tuning by practitioners. Thus, the performance of these prior heuristic-based local learning methods is limited. This work targets the mainstream bias problem by proposing an end-to-end adaptive local learning framework to automatically and adaptively learn customized local models for different users, overcoming the limitations of heuristic-based methods to mitigate mainstream bias.

\section{Conclusion}
In this study, we aim to address the mainstream bias in recommender systems that niche users who possess special and minority interests receive overly low utility from recommendation models. We identify two root causes of this bias: the discrepancy modeling problem and the unsynchronized learning problem. Toward debiasing, we devise an end-to-end adaptive local learning framework: we first propose a loss-driven Mixture-of-Experts module to counteract the discrepancy modeling problem, and then we develop an adaptive weight module to fight against the unsynchronized learning problem. Extensive experiments show the outstanding performance of our proposed method on both niche and mainstream users and overall performance compared to SOTA alternatives. 

\subsubsection*{Acknowledgements.}
This research was funded in part by 4-VA, a collaborative partnership for advancing the Commonwealth of Virginia.

%
%
%
\bibliographystyle{splncs04}
\bibliography{mybibliography}
%




\end{document}